# Car-following behavior of connected vehicles in a mixed traffic flow: modeling and stability analysis*

Lin Liu[1, 2, 3], Chunyuan Li[1, 2], Yongfu Li[1, 2], Srinivas Peeta[3], Lei Lin[3]

*Abstract*— Vehicle-to-vehicle communications can change the driving behavior of drivers significantly by providing them rich information on downstream traffic flow conditions. This study seeks to model the varying car-following behaviors involving connected vehicles and human-driven vehicles in a mixed traffic flow. A revised car-following model is developed using an intelligent driver model to capture drivers' perceptions of their preceding traffic conditions through vehicle-to-vehicle communications. Stability analysis of the mixed traffic flow is conducted for a specific case. Numerical results show that the stabile region is apparently enlarged compared with the IDM model.

## I. Introduction

Connected vehicle technologies in the recent years have seen a rapid growth and received tremendous interests from academics, industries and government agencies. US Department of Transportation has delivered a notice of Proposed Rulemaking about V2V technology in 2016 and will deliver a regulation mandating Vehicle-to-vehicle (V2V) technology this year. V2V technology will begin to be used in new car production from 2019 and will be included on 100% of new car production from 2021. This shows that there will be a long transition time of a mixed traffic flow with human-driven vehicles and connected vehicles in the near future. V2V wireless communications can provide an opportunity to create an internet of vehicles where individual vehicles can exchange movement information such as vehicle location, velocity and acceleration with each other. In this context of driving, drivers can not only take into account the behavior of the immediate leader, but also, they can get rich information on downstream traffic flow conditions. Consequently, all aspects of drivers' decision making, both strategic and operational decisions, would be impacted and generally enhanced. At the operational level, V2V communication can help drivers make safer and more reliable decisions about acceleration choice for a more stable car-following behavior.

From the theoretical standpoint, it's challenging to model the car-following behavior of connected vehicles under a mixed traffic flow with considering the effect of connectivity. Car-following behavior has been studied extensively in the literature and several models with different levels of complexity have been introduced to capture the underlying process of acceleration decision making, such as Pipes, Forbes, General Motors and Optimal velocity model [1-7]. However, most of these previous studies focused on the car-following behavior of human-driven vehicles without presenting the role of connectivity, although some studies have introduced the spatial anticipation model [8], in which the driver of the following vehicle is regarded that he can perceive the information from multi-preceding vehicles. As for connected vehicles with the capability of communication, the acceleration choice is usually modeled using a traditional car-following model-IDM (Intelligent Driver Model) without considering the information provided by multiple preceding connected vehicles [9]. What's more, some efforts focused on specific applications of connected and autonomous vehicle (CAV) technologies in pure CAV driving environment, such as Cooperative Adaptive Cruise Control or Automated Highway Systems, from the perspective of control without considering the interaction of different kinds of vehicles [10-11]. The capabilities of these modeling are limited to in a mixed environment where only a portion of vehicles is equipped with the essential communication tools. Therefore, there is a need for a new model to capture the car-following behaviors of vehicles in a heterogeneous platoon including human-driven vehicles and connected vehicles.

The remainder of this paper is organized as follows: in Section II, the proposed car-following model is formulated in terms of the acceleration function. In Section III, the stability of this model for a specific case is analyzed, and the comparison between this model and IDM is discussed. A conclusion is given in the final Section IV.

*Research is supported by Chongqing Research Program of Basic Research and Frontier Technology (cstc2017jcyjAX0402), Scientific and Technological Research Program of Chongqing Municipal Education Commission (KJ1600416) and Natural Science Funds of CQUPT (A2013-27).

Lin Liu is with College of Automation, Industrial IoT Collaborative Innovation Center, College of Automation, Chongqing University of Posts and Telecommunications, Chongqing 400065, China (corresponding author to provide phone: 13752944485; e-mail: liulin@cqupt.edu.cn).
Chunyuan Li and Yongfu Li are with College of Automation, Industrial IoT Collaborative Innovation Center, College of Automation, Chongqing University of Posts and Telecommunications, Chongqing 400065, China (e-mail: mmlicy@126.com, laf1212@163.com).
Srinivas Peeta and Lei Lin are with Lyles School of Civil Engineering, Purdue University, West Lafayette, IN 47907 United States of America (e-mail: peeta@purdue.edu; lin954@purdue.edu).

## II. Modeling behavior of connected vehicles

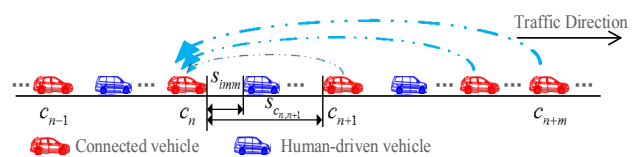

Figure 1. Car following in the mixed traffic flow

Under the mixed traffic flow with connected vehicles and human-driven vehicles, there are different numbers of human-driven vehicles between two adjacent connected vehicles, such as zero, one, two or more. All the connected vehicles can communicate with each other wirelessly. In the context of driving, as shown in Fig.1, a human driver of connected vehicle $c_n$ not only takes into account the behavior of the immediate leader, but also receives the movement information of several connected vehicles ahead. For car-following model, we just consider that the following vehicle receives the information from all the preceding connected vehicles in the communication range.

Considering the effects of the preceding connected vehicles, we split the acceleration into two parts, which are the traditional car following of the immediate preceding vehicle $imm$ and the effects of $m$ connected vehicles in front. $m$ is the maximum connected vehicles in the communication range. We supposed that $x_{c_n}$, $v_{c_n}$ and $a_{c_n}$ are the position, velocity and acceleration of vehicle $c_n$, respectively. The mathematical description of this idea is following:

$$a_{c_n} = a_{imm} + a_{cv} = \underbrace{g(s_{imm}, v_{c_n}, \Delta v_{imm})}_{car-following} + \underbrace{\sum_{k=1}^{m} f(s_{c_{n,n+k}}, a_{c_{n+k}}, \Delta v_{c_{n,n+k}})}_{CVs'-effects} \quad (1)$$

Where $s_{imm}$ and $\Delta v_{imm}$ are the headway and velocity difference between the immediate preceding vehicle $imm$ and the following vehicle $c_n$, and $s_{imm} = x_{imm} - x_{c_n} - l_{imm}$, $\Delta v_{imm} = v_{c_n} - v_{imm}$. $l$ is the length of a vehicle. $s_{c_{n,n+k}}$ and $\Delta v_{c_{n,n+k}}$ are the headway and velocity difference between the connected vehicle $c_n$ and $c_{n+k}$, and $s_{c_{n,n+k}} = x_{c_{n+k}} - x_{c_n} - l_{c_{n+k}}$, $\Delta v_{c_{n,n+k}} = v_{c_n} - v_{c_{n+k}}$.

The driving behavior (e.g. headway, velocity, acceleration) of the immediate preceding vehicle is perceived by the driver of the following vehicle and this can be captured using a traditional car following model. Here we choose the IDM (Intelligent Driver Model) proposed by Treiber [13]. So the effect of the immediate preceding vehicle on the following vehicle can be shown as follows.

$$a_{imm} = g(s_{imm}, v_{c_n}, \Delta v_{imm}) = a_0 \left[ 1 - \left(\frac{v_{c_n}}{v_0}\right)^\delta - \left(\frac{s^*(v_{c_n}, \Delta v_{imm})}{s_{imm}}\right)^2 \right] \quad (2)$$

The IDM acceleration consists of a free acceleration $a^{free} = a_0 \left[ 1 - (v_{c_n}/v_0)^\delta \right]$ for approaching the desired velocity $v_0$ with an acceleration slightly below the maximum acceleration $a_0$, and the braking interaction $a^{int} = a_0 (s^*/s_{imm})$, where the actual gap $s_{imm}$ is compared with the desired minimum gap $s^*$ that is defined as below. $\delta$ is a free acceleration exponent.

$$s^*(v_{c_n}, \Delta v_{imm}) = s_0 + v_{c_n} T + \frac{v_{c_n} \Delta v_{imm}}{2\sqrt{a_0 b_0}} \quad (3)$$

Which is specified by the sum of the jam distance $s_0$, the velocity-dependent safety distance $v_{c_n} T$ corresponding to the safe time headway $T$, and a dynamic part. The dynamic part implements an accident-free intelligent braking strategy that, in nearly all situations, limits braking decelerations to the comfortable deceleration $b_0$.

These multiple connected vehicles ahead will communicate the driving behavior of themselves with the following vehicle and this information will affect the behavior of the following vehicle, which will be considered from the perspective of safe movement. Therefore, the following vehicle can update its acceleration as a function of its current velocity and the knowledge of the velocity and headways of its preceding connected vehicles within the communication range. The idea is then to integrate all the information with the weighting coefficients as Eq.(4).

$$a_{cv} = \sum_{k=1}^{m} f(s_{c_{n,n+k}}, a_{c_{n+k}}, \Delta v_{c_{n,n+k}})$$
$$= -K_v \sum_{k=1}^{m} \alpha_k \left(s_{c_{n,n+k}}\right) \Delta v_{c_{n,n+k}} + K_a \sum_{k=1}^{m} \beta_k \left(s_{c_{n,n+k}}\right) a_{c_{n+k}} \quad (4)$$

Where $K_v$ and $K_a$ are the different constant sensitivity coefficients. $K_v \geq 0, K \geq 0$ and $K_v, K_a \in [0,1]$. $\alpha_k$ and $\beta_k$ are the weighted parameters, which are the function of the headway. $\alpha_k \geq 0$ and $\beta_k \geq 0$. For convenience, we suppose $\alpha_k = \beta_k = \omega_k$.

III. STABILITY ANALYSIS

Based on the different market penetration rates of connected vehicles, there will be lots of configurations for the mixed platoon including connected vehicles and human-driven vehicles. In this paper, we just consider one specific case to study the string stability. We assume that the market penetration rate of connected vehicle is 35%. That means there will be two human-driven vehicles in between two adjacent connected vehicles which is illustrated in Fig.2.

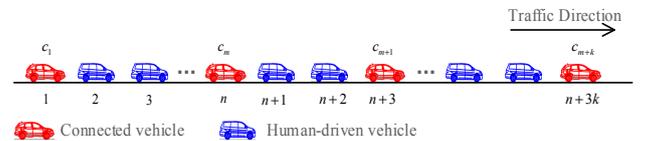

Figure 2. A specific configuration of platoon

We assume that there are $M$ connected vehicles in front of connected vehicle $c_m$. As for this specific case, the model can be written as follows:

$$a_{c_m} = g(s_{imm}, v_{c_n}, \Delta v_{imm}) + \sum_{k=1}^{M} f(s_{c_m, c_{m+k}}, a_{c_{m+k}}, \Delta v_{c_m, c_{m+k}}) \quad (5)$$

The sequence number of connected vehicles has a relationship with the sequence number of the heterogeneous platoon, which is compatible with this condition $n = 3m - 2$. As a result, the acceleration of vehicle $n$ in the platoon can be rewritten as below:

$$a_n = g(s_{n+1}, v_n, \Delta v_{n+1}) + \sum_{k=1}^{M} f(s_{n,n+3k}, a_{n+3k}, \Delta v_{n,n+3k})$$
$$= a_0 \left[ 1 - \left(\frac{v_n}{v_0}\right)^\delta - \left(\frac{s^*(v_n, \Delta v_{n+1})}{s_{n+1}}\right)^2 \right]$$
$$- K_v \sum_{k=1}^{M} \alpha_k (s_{n,n+3k}) \Delta v_{n,n+3k} + K_a \sum_{k=1}^{M} \beta_k (s_{n,n+3k}) a_{n+3k} \quad (6)$$

Equilibrium state is reached when $\Delta v_n = 0$ and $a_n = 0$. The steady-state position, speed and headway of vehicle $n$ are $x_n^0(t), v_n^0$ and $S_n^0$, respectively. The position of each vehicle $n$ in the stabile traffic flow at time t is:

$$x_n^0(t) = nS_n^0 + v_n^0 t, \quad n=1,2,3...,N \quad (7)$$

Small position perturbations $y_n(t)$ is added around an equilibrium state $x_n^0(t)$, so the position of vehicle $n$ at time $t$ is:

$$x_n(t) = x_n^0(t) + y_n(t) \quad (8)$$

Eq.(8) can be rewritten as the following formation:

$$y_n(t) = x_n(t) - x_n^0(t) \quad (9)$$

And $\Delta x_n(t) = \Delta y_n(t) + S_n^0$, $\dot{y}_n(t) = \dot{x}_n(t) - v_n^0$, $\ddot{y}_n(t) = \ddot{x}_n(t)$, substitute Eq.(9) into the Eq.(6) and using the Taylor expansion, it will deduce:

$$\begin{aligned}
&\dot{y}_n(t+2T') - \dot{y}_n(t+T') \\
&= T'g_1(y_{n+1}(t) - y_n(t)) \\
&+ g_2(y_n(t+T') - y_n(t)) \\
&+ g_3(y_n(t+T') - y_n(t) - y_{n+1}(t+T') + y_{n+1}(t)) \\
&- f_4 \sum_k \omega_k (y_n(t+T') - y_n(t) - y_{n+3k}(t+T') + y_{n+3k}(t)) \\
&+ f_5 \sum_k \omega_k (\dot{y}_{n+3k}(t+2T') - \dot{y}_{n+3k}(t+T'))
\end{aligned} \quad (10)$$

Where $T'$ is the reaction time of the driver. In order to simplify the complexity of format, we have used the partial derivatives $g_i$, $i = 1, 2, 3$ of $g$ to $S_n$, $v_n$, $\Delta v_n$, and the partial derivatives $f_i$, $i = 4, 5$ of $f$ to $\Delta v_n$, $a_n$, respectively.

Specifically, $g_1 = \partial g / \partial s_n (S^0, v^0, 0)$, $g_2 = \partial g / \partial v_n (S^0, v^0, 0)$, $g_3 = \partial g / \partial \Delta v_n (S^0, v^0, 0)$, $f_4 = \partial f / \partial \Delta v_n (S^0, v^0, 0)$ and $f_5 = \partial f / \partial a_n (S^0, v^0, 0)$.

Set $y_n(t) = \exp(i\alpha n + zt)$ and according to Fourier transform, we can obtain:

$$(e^{T'z} - 1) \begin{pmatrix} ze^{T'z} - g_2 + g_3(e^{i\alpha} - 1) \\ + f_4 \sum_k \omega_k (e^{i\alpha 3k} - 1) \\ - f_5 \sum_k \omega_k (ze^{i\alpha 3k} e^{T'z}) \end{pmatrix} = T'g_1(e^{i\alpha} - 1) \quad (11)$$

Let $z = z_1(i\alpha) + z_2(i\alpha)^2 + \cdots$, and expand it to the second term of $(i\alpha)$, we will have the expression of $z_1$ and $z_2$. When $z_2 > 0$, the model is stable and the stability condition is:

$$g_1 - \frac{1}{2}g_2^2 - g_1g_2T' - g_2g_3 + f_4 \sum_j \omega_j - g_2 f_5 \sum_j \omega_j < 0 \quad (12)$$

If Eq.(12) is satisfied, the proposed model is stable. That is, the perturbation will decay as it propagates to the upstream traffic. Otherwise, the perturbation will lead to a collision or a traffic congestion.

Typical values of the proposed model parameters are used as shown in Table 1.

TABLE I. ACCELERATION MODEL PARAMETERS AND THEIR VALUES

| Parameters | Typical values |
|---|---|
| Desired Velocity $v_0$ | 33.3 m/s |
| Safe Time Headway $T$ | 1.6 s |
| Maximum Acceleration $a_0$ | 0.73 m/s$^2$ |
| Comfortable Deceleration $b_0$ | 1.67 m/s$^2$ |
| Acceleration Exponent $\delta$ | 4 |
| Linear Jam Distance $s_0$ | 2 m |
| Reaction time $T'$ | 1s |

Note that the partial derivatives of IDM are functions of vehicle speed and gap at equilibrium; therefore, the following relationship between speed and equilibrium gap is used to simplify the stability analysis [14].

$$s_e(v_n) = \frac{s_0 + v_n T}{\sqrt{1 - \left(\frac{v_n}{v_0}\right)^\delta}} \quad (13)$$

Based on the stable condition Eq.(12), the stable diagram is plotted in Fig.3 with maximum acceleration $a_m$ and desired time headway $T_d$.

From the numerical results in Fig.3, the region over the critical curve is the stable region; while the remainder is the unstable region. It reveals that the stabile region of the proposed car-following model is larger than that of IDM. The reason is that the proposed model takes into account the effects of the preceding connected vehicles within the communication range. With the increase of $M$, that is, further considering the mere preceding vehicles' information, the stable region will be enlarged, and up to tend to a fixed area. The increase of stable region of traffic flow indicates that vehicles can move faster than before at the same time headway, which is meant to suppress traffic jams effectively.

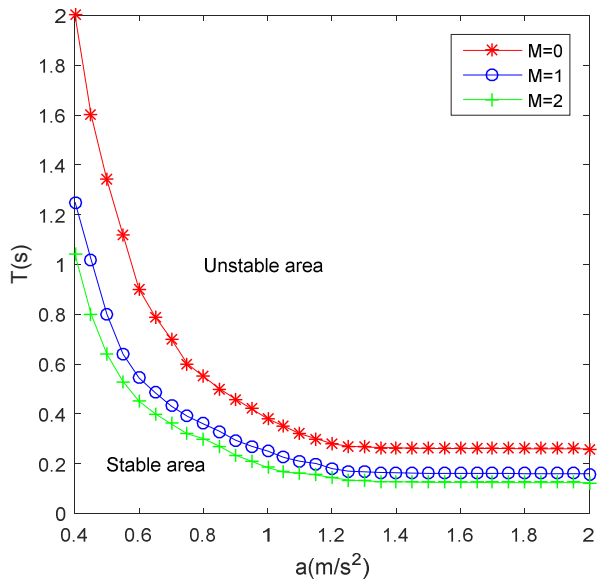

Figure 3. The stable diagram of the proposed model with different numbers of preceding connected vehicles

## IV. Conclusion

An extended IDM model is proposed to capture the car-following behavior of connected vehicles under a heterogeneous platoon and stability analysis is done for a special case. Numerical results show that the extended IDM is more stable than IDM, which is benefit for easing the traffic jam. For future research directions, we will explore string stability for general cases and calibrate the proposed model.

## Acknowledgment

The authors would like to thank Dr. Siyuan Gong for his insightful comments and Dr. Jian Wang who has improved the quality of the paper.